\documentstyle[12pt]{article}
\newcommand{\be}{\begin{equation}}
\newcommand{\ee}{\end{equation}}
\newcommand{\ba}{\begin{eqnarray}}
\newcommand{\ea}{\end{eqnarray}}

\begin{document}
\begin{center}
 {\bf\Large{Multi Hamilton-Jacobi quantization of
 O(3) nonlinear sigma  model }}
\end{center}
\begin{center} {\bf Dumitru Baleanu}\footnote[1]{ On leave of absence from
Institute of Space Sciences, P.O BOX, MG-23, R 76900
Magurele-Bucharest, Romania,
E-mail: dumitru@cankaya.edu.tr}
and
{\bf Yurdahan G\"uler}\footnote[2]{E-Mail:~~yurdahan@cankaya.edu.tr}
\end{center}
\begin{center}
Department of Mathematics and Computer Sciences, Faculty of Arts
and Sciences, Cankaya University-06530, Ankara , Turkey
\end{center}
\begin{abstract}
The $O(3)$ non-linear sigma model
is investigated using multi \\
Hamilton-Jacobi formalism.
The integrability conditions are investigated and the results
are in agreement with  those obtained by Dirac's  method.
By choosing an adequate extension of phase space we describe the
transformed system by a set of three Hamilton-Jacobi equations and
calculate the corresponding action.
\end{abstract}
\newpage

 \section {Introduction}

 The importance of $O(3)$ nonlinear sigma model lies  in theoretical
and phenomenological basis \cite{unu, doi, trei}.
It is a fact that this theory describes classical (anti) ferromagnetic spin systems
at their critical points in the Euclidean space, while in the Minkowski one
it delineates the long wavelenght limit of quantum antiferromagnets.
On the other hand the model exhibits solitons, Hopf instantons and novel spin and statistics in
2+1 space-time dimensions with inclusion of the Chern-Simon term \cite{trei,patru}.

 Recently, the quantization of this model was investigated \cite{k0} by
improved Batalin-Fradkin-Tyutin method \cite{kim}.
An alternative method of quantization of the systems with constraints
was initiated in \cite{g5, g6} and it is based on the
 $Carath{\acute e}odory$'s equivalent Lagrangians method \cite{car}.

  Recently, the singular systems
with higher order Lagrangians ,the systems which have  elements
of the Berezin algebra \cite{p11, p12, p13},
the quantization of Proca's model \cite{gu11} ,
the non-relativistic particle on a curved space \cite{gb7} as well as  the supersymmetric
 quantum mechanics \cite{gb8} in
Witten's version \cite{wit} were investigated using this method.

One of the most interesting application  of the formalism is the case
when the systems admit second class constraints \cite{dirac, gomis}
simply because the corresponding system of equations is not integrable \cite{gomis}.
Is it possible to quantize the system using multi Hamilton-Jacobi formalism
in this case?
The answer is positive if we transform the system such that it becomes
completely integrable and in addition the form of the Hamiltonians are
suitable for application of multi Hamilton-Jacobi equations.

On the other hand  the multi Hamilton-Jacobi formalism for fields
requires a special attention  when all the
Hamiltonians are densities and, as a consequence, the surface
terms \cite{henneaux} play an important role in the process of quantization.

For these reasons the quantization of the
$O(3)$ nonlinear sigma model using multi Hamilton-Jacobi formulation is interesting
to investigate.

The plan of the paper is the following:

In sec. 2 the multi Hamilton-Jacobi formalism for fields  is presented.
In sect. 3 the path integral quantization of O(3) model is analyzed.
Sec. 4 contains our conclusions.

\section{Multi Hamilton-Jacobi formalism for fields}

Let us assume that we  start with a singular Lagrangian density having
the  Hessian matrix of rank n-r .
Using the  $Carath{\acute e}odory$'s equivalent Lagrangian method
we find the  following Hamiltonian densities

\be\label{doi} H_{\alpha}^{'}=H_{\alpha}(t_{\beta},q_{a},p_{a})
+p_{\alpha}, \ee where $\alpha,\beta=n-r +1,\cdots,n$,$a=1,\cdots
n-r$. The canonical Hamiltonian $H_0$ is defined as

\be\label{unu} H_{0}=-L(t,q_{i},{\dot q_{\nu}},{\dot q_{a}=w_{a}})
+p_{a}w_{a} + {\dot
q_{\mu}}p_{\mu}\mid_{p_{\nu}=-H_{\nu}},\nu=0,n-r+1,\cdots,n. \ee
and by contruction is  independent of $\dot q_{\mu}$. Here $\dot
q_{a}={dq_{a}\over d\tau}$,where $\tau$ is a parameter.
 The equations of motion are obtained as total differential equations
in many variables as follows

\ba\label{(pq)}
&dq_{a}&= {\delta H_{\alpha}^{'}\over\delta
p_{a}}dt_{\alpha},
dp_{a}=-{\delta H_{\alpha}^{'}\over\delta
q_{a}}dt_{\alpha},\cr
&dp_{\mu}&=-{\delta H_{\alpha}^{'}\over\delta
t_{\mu}}dt_{\alpha}, \mu=1,\cdots, r, \ea

\be\label{(z)} z=\int{(-H_{\alpha} + p_{a}{\delta
H_{\alpha}^{'}\over\delta p_{a}})}dt_{\alpha}, \ee where
$z=S(t_{\alpha},q_{a})$ and ${{\delta H}\over \delta x}$
represents the variation  of H with respect to x.
Since
the equations of motion are total differential equations the
integrability conditions plays an important role.
More exactly  eqs.
(\ref{(pq)},\ref{(z)} )are integrable iff $dH_{\alpha}{'}=0$ .
If the variations of
$H_{\alpha}{'}$ are not zero then  additional constraints may arise.
Thus, we may have  Hamiltonian densities  other than (\ref{doi}).
The essence of the formalism  is to
express all Hamiltonian densities  in the form (\ref{doi}) and  (\ref{unu}).

\section{Multi Hamilton-Jacobi treatment of O(3) nonlinear sigma  model}
The model is described by the Lagrangian
\be\label{lag}
L=\int{d^2x}\left[{1\over 2f}(\partial_{\mu}n^{a}\partial^{\mu}n^{a})-
{\dot\lambda} (n^{a}n^{a}-1)\right],
\ee
 where the metric has signature  $(+,-,-)$, $n^{a}(a=1,2,3)$ is a multiplet of three real scalar field with
 a constraint
 \be\label{con}
 n^{a}n^{a}-1=0
 \ee

 and  ${\dot \lambda}={d\lambda\over d\tau}$ is a Lagrange multiplier.

In multi Hamilton-Jacobi formalism
we have two Hamiltonian densities which corresponds to (\ref{lag})
\ba\label{ham1}
&H_{0}^{'}&=\int{d^2x}[p_{0}+{f\over 2}\pi^{a}\pi^{a} +
{1\over 2f}(\partial_{i}n^{a})(\partial_{i}n^{a})],i=1,2\cr
&H_{1}^{'}&=\int{(\pi_{\lambda}+n^{a}n^{a}-1)}d^{2}x.
\ea

Here $\pi_{a}={1\over f}{\dot n^{a}}$ and $\pi_{\lambda}$
represents the momentum conjugate to $\lambda$.

Integrability condition $dH_{2}^{'}=0$ gives
\be\label{con2}
n^{a}\pi^{a}=0.
\ee
The variation of (\ref{con2}) determines $\lambda$
as the solution of the differential equation
\be\label{lala}
{\dot\lambda}=-{1\over n^{a}n^{a}}({f\over 2}\pi^{a}\pi^{a} +{1\over 2f}n^{a}
{\partial_{i}^{2}n^{a}}).
\ee
In conclusion we have three Hamiltonian densities for our model
\ba\label{ham2}
&H_{0}^{'}&=\int{d^2x}[ p_{0}+{f\over 2}\pi^{a}\pi^{a} +{1\over 2f}(\partial_{i}n^{a})(\partial_{i}n^{a})
],\cr
&H_{1}^{'}&=\int{(\pi_{\lambda}+ n^{a}n^{a}-1)}d^2x, H_{2}^{'}=\int n^{a}\pi^{a}d^2x.
\ea
We would like to stress on the fact that $\pi_{\lambda}$ is a constant
and
the results  (\ref{ham2}) are in agreement with Dirac's analysis \cite{dirac}.
The next step is to quantize the model by transforming the Hamiltonian
densities (\ref{ham2}) to be in involution and calculate the corresponding characteristic.

The essence of  Batalin-Fradkin-Tyutin  formalism  is to  enlarge the phase space  with
some extra variables such that the modified canonical Hamiltonian and modified
second class constraints to be in involution (for more details see Refs. \cite{fad}).
Recently , a formalism , which is called improved Batalin-Fradkin-Tyutin has been proposed
(see Refs. \cite{kim} and the references therein)
Let us consider a constrained system having only second class constraints
$\Phi_{\alpha}$.
This formalism contains two steps, the first step consists in
finding a set of constraints in involution and the second step deals with
transformation of all fields and corresponding momenta in such a way that they
are in involution with the transformed constraints.

In our  specific problem, since we have only two second class constraints,
we need only two extra fields $\theta, \pi_{\theta}$ to start with.
On the other hand we have to find a set of Hamiltonians in involution
and in the form (\ref{doi}). To reach this objective we will exploit the fact that
the commutation relations of the extra variables in Batalin-Fradkin-Tyutin
formalism  are not uniquely defined , so we will try to find a proper choice
such that the transformed Hamiltonians  to be in form (\ref{doi}).
We assume that  $\pi_{\theta}$ is the canonical conjugate momenta of $\theta$ .
Next step is to make the Hamiltonians $H_{1}^{'}$ and $H_{2}^{'}$
in involution.

We are looking to find  $H_{i}^{''}$ as
\be
H_{i}^{''}=\sum_{n=0}^{\infty}\Phi_{i}^{n}, i=1, 2
\ee

fulfiling the boundary conditions $\Phi_{i}^{0}=H_{i}^{'}$ such that
\be\label{haha}
\{H_{i}^{''}(x),H_{j}^{''}(y)\}=0.
\ee
Here $\Phi_{i}^{n}$ are polynomials in $\theta$ and $\pi_{\theta}$ .

Due to linearity in the auxiliary fields we can make the following   ansatz
\be\label{xas}
 \Phi_{A}^{1}=\int{d^2y}X_{AB}(x,y)\Gamma^{B}(y),
\ee
where $\Gamma^{B}$ represents the set $(\theta,\pi_{\theta})$\cite{fad}
and $X_{AB}(x,y)$ is a matrix to be determined.

{}From  (\ref{haha}) and (\ref{xas}) we  find the equations corresponding to
$X_{AB}$ as
\be\label{xx}
\Delta_{AB} + X_{AC}\omega^{CD}X_{BD}=0,
\ee

where $\Delta_{AB}=2\epsilon^{AB}n^a n^a\delta(x-y)$ and
$\epsilon^{12} =-\epsilon^{21}=1$.
 The solution of  (\ref{xas}) is

\be\label{xxu}
X_{AB}={\pmatrix{2&0\cr
0&n^an^a\cr}}\delta(x-y).
\ee

Using (\ref{xas}) and (\ref{xxu})
we find the  following strongly involutive  Hamiltonians
\be
H_{2}^{''}= \int d^{2}{x}( {\pi_{\lambda}+n^{a}n^{a}-1+2\pi_{\theta}}),
H_{3}^{''}=\int d^2{x}({n^{a}\pi^{a}+n^{a}n^{a}\theta}).
\ee

The next step is to transform  $H_{0}^{'}$ such that it is in involution
with $H_{2}^{''}$ and $H_{3}^{''}$.
Denoting $\Sigma=(n^a,\pi^a)$ we would like to make  a
transformation such that $\{n^a,H_{2}^{'}\}=\{\pi^a,H_{2}^{'}\}=0$
and as a consequence
$\{H_{0}^{'},H_{2}^{'}\}=\{H_{0}^{'},H_{3}^{'}\}=0$ \cite{kim}.
Expressing ${\hat\Sigma}$ to be the set $({\hat n^a},{\hat \pi^a})$
we can reach  this objective making the assumption

\be
{\hat\Sigma}= \Sigma + \sum_{n=1}^{\infty}{\hat\Sigma}^{(n)},
{\hat\Sigma}^{(n)}\sim \Gamma^{(n)},
\ee
where the $(n+1)$-th order of iteration has the form \cite{kim}
\be
{\hat\Sigma}^{(n+1)}=-{1\over n+1}\int{d^2xd^2yd^2z \Gamma^{A}(x)\omega_{AB}X^{BC}(y,z)G_{C}^{(n)}(z)}
\ee
with
\ba
&G_{A}^{(n)}(x)&=\sum_{m=0}^{n}\{H_{A}^{'(n-m)},{\hat\Sigma^{(m)}}\}_{(\Sigma)} +
\sum_{m=0}^{n-2}\{\{H_{A}^{'(n-m)},{\hat\Sigma^{(m+2)}}\}\}_{(\Gamma)}\cr
& +&
\{H_{A}^{'(n+1)},{\hat\Sigma^{(1)}}\}_{(\Gamma)}.
\ea
The calculations gives the following expressions for the set $\hat\Sigma$
\ba\label{tti}
&{\hat n^a}&=n^{a}\left(1-\sum_{n=1}^{\infty}{(-1)^{n}\pi_{\theta}^{n}(2n-3)!!\over (n^an^a)^n n!}\right),\cr
&{\hat \pi^a}&=(\pi^{a}+n^{a}\theta)\left(1+\sum_{n=1}^{\infty}{(-1)^{n}\pi_{\theta}^{n} (2n-1)!!\over (n^an^a)^n n!} \right).
\ea
\be\label{hami}
H_{0}^{''}=\int{d^2x}\left( p_{0}+{f\over 2}{\hat{\pi^{a}}}{\hat{\pi^{a}}} +
{1\over 2f}\partial_{i}{\hat n^{a}}\partial_{i}{\hat n^{a}}\right).
\ee
Taking into account (\ref{tti}) and the form of series expansion
of $f(x)={1\over 1+2x}$ around $x=0$ and using the conformal map condition
$n^a{\partial_{i}n^a}=0$ we reexpress (\ref{hami}) in a simpler form
\ba\label{n1ham}
&H_{0}^{'''}&=\int{d^2x}[ p_{0}+{f\over 2}(\pi^{a}+n^{a}\theta)(\pi^{a}+n^a\theta){n^b n^b\over{n^b n^b+ 2\pi_{\theta}}} \cr
 & + & {1\over 2f}\partial_{i}{n^{a}}\partial_{i}{ n^{a}}{n^b n^b +2\pi_{\theta}\over n^b n^b}].
\ea
Even if the Hamiltonian densities  $H_{0}^{'''},H_{1}^{'''},H_{2}^{'''}$ are
in involution $H_{1}^{'''}$  and $H_{2}^{'''} $ aren't in the form given
by (\ref{doi}) and
then we can not describe, yet, the transformed system
with three Hamilton-Jacobi  equations.
For these reasons we must modify the form
of $H_{1}^{'''}$ and $H_{2}^{'''}$.
Dividing $H_{1}^{'''}$ and $H_{2}^{'''}$ by 2 and respectivelly by $n^{1}$
we get

\ba\label{nham}
&H_{1}^{'''}&= \int d^{2}{x}\left( \pi_{\theta}+{{{\pi_{\lambda}+n^{a}n^{a}-1}\over 2}}\right),\cr
&H_{2}^{'''}&= \int d^2{x}\left(\pi^{1}+{{n^{2}\pi^{2}+n^{3}\pi^{3}+n^{a}n^{a}\theta\over n^{1}}}\right)
\ea
and by inspection we conclude that $H_{0}^{'''},H_{1}^{'''},H_{2}^{'''}$
are weakly involutive.
>From (\ref{n1ham}) and (\ref{nham}) and taking into account (\ref{(z)}) we find
the form of the action as
\ba
&z&=\int d^2xd\tau[{f\over 2}((\pi^{2})^2  + (\pi^{3})^2 -(n^{2}\theta)^2
+ {{-n^{a}n^{a}+1}\over 2}{\dot\theta}
-(n^{3}\theta)^2  \cr
&-&(\pi^{1}+n^{1}\theta)^{2} ){n^b n^b\over{n^b n^b+ 2\pi_{\theta}}}
 -{1\over 2f}\partial_{i}{n^{a}}\partial_{i}{ n^{a}}{n^b n^b +2\pi_{\theta}\over n^b n^b}
-{n^{a}n^{a}\theta\over n^{1}}{\dot\pi_{1}}].
\ea

\section{Conclusion}

 The multi Hamilton- Jacobi formalism has the unique attribute of exhibiting
 the classical analog of the quantum state.

The treatment of the second class constrained systems is problematic for
multi Hamilton-Jacobi procedure. In fact two types of problems arises,
the first and the most important one is that the initial system is not
integrable and the second one is related to the form of the constraints.

To solve these problems we have to transform the system by enlarging
or reducing the initial phase space such that the transformed Hamiltonians
become in involution. On the other hand we have to preserve the form of the
Hamiltonians such that to be suitable for application of $Carath{\acute e}odory$'s
equivalent Lagrangian method.

In this paper we investigate both from classical and quantum point of view,
the $O(3)$ nonlinear sigma model using multi Hamilton-Jacobi formulation.
>From the consistency conditions we find  the parameter $\lambda$ which is
in agreement with  Dirac's procedure.

To quantize the system we extend the phase-space using
improved Batalin-Fradkin-Tyutin procedure choosing a specific form of
commutation relation between extra fields. After finding the Hamiltonians
densities in involution and in the form  (\ref{doi})
we calculate the action.

\section {Acknowledgments}
DB would like to thank M. Henneaux for the valuable
communication.
This paper is partially supported
by the Scientific and Technical Research Council of Turkey.

\end{document}